\documentclass{cpbtex}

\usepackage{amsfonts}
\usepackage{graphicx,,booktabs}
\usepackage{amssymb}
\usepackage{amsmath,txfonts}
\usepackage{graphicx}
\usepackage{dcolumn}
\usepackage{color}
\usepackage{bm}

\begin{document}

\title{Nonlinear signal transduction network with multistate\thanks{Project supported by the National Natural Science Foundation of China (Grants No. 11675228) and China postdoctoral Science Foundation (Grants No. 2015M572662XB).}}


\author{Han-Yu Jiang, Jun He\thanks{Corresponding author. E-mail:~junhe@njnu.edu.cn}\\
School of Physics and Technology, Nanjing Normal University,
Nanjing, Jiangsu 210097, China}   


\date{\today}
\maketitle

\begin{abstract}
Signal transduction is  an important and basic mechanism to cell life activities. The stochastic state transition of receptor induces the release of signaling molecular, which triggers the state transition of other receptors. It constructs a nonlinear sigaling  network,  and leads to robust switchlike properties which are critical to  biological function. Network architectures and  state transitions of receptor affect the performance of this biological network.  In this work, we perform a study of nonlinear signaling on biological polymorphic network by analyzing network dynamics of the Ca$^{2+}$ induced Ca$^{2+}$ release mechanism, where fast and slow processes are involved and the receptor has four conformational states.   Three types of networks, Erd\"os-R\'enyi network,  Watts-Strogatz network and BaraB\'asi-Albert network, are considered with different parameters. The dynamics of the biological networks exhibit different patterns at different time scales. At short time scale,  the second open state is essential to reproduce the quasi-bistable regime, which emerges at a critical  strength of connection for all three states involved in the fast processes and disappears at another critical point.   The pattern at short time scale is not sensitive to the network architecture. At long time scale,  only monostable regime is observed, and  difference of   network architectures affects the results more seriously. Our finding identifies features of nonlinear signaling networks with multistate that may underlie their biological function.
\end{abstract}

\textbf{Keywords:} Signal transduction, Biological network with multistate, CICR, Nonlinear signaling

\textbf{PACS:} 87.18.Mp,  87.16.Xa, 87.17.Aa

\section{Introduction}

The signal transduction is important to a variety of intracellular and intercellular processes~\cite{Kalckar1991}. The signal propagates between different cellular components  and communicates between biological cells. It usually mediates between two receptors by  signaling molecule, which changes the conformational state of receptor protein. Generally,  the response of receptor is typically a nonlinear function of the local concentration of  signaling molecule~\cite{Fang2019,Braichenko2018,Hernandez-Hernandez2017}. In this way, signal transduction systems can perform robust switchlike operations~\cite{Kholodenko2006,Fang2019}.  The state transition of the receptors  triggers or inhibits  the release of signaling molecule, which then affects other receptors' state. With such mechanism, the signal propagates on the network of intracellular and intercellular receptors. Given that signal transduction involves a large number of receptors with complex connections by  signaling molecule, it is natural to apply the network science to understand their dynamics~\cite{Albert2004,Tong2004,Barabasi2004,Barabasi20016}.   The network sciences has been widely practiced in different research fields~\cite{Rafo2021,Wu2020,Li2019,Viguerie2021,Wang2015}.   To keep and regulate the life activity, there exist a variety of signaling networks, such as calcium signaling network, protein-protein interaction networks, and excitation-contraction coupling in muscle tissue~\cite{Bers2002,Li2004,Berridge1998}.  Up to now, much efforts have been paid in this direction and some key insights have been achieved using such interdisciplinary approach~\cite{Gosak2018}.

In the signal  transduction network, the receptors are taken as the nodes which are connected by  signaling molecule.  Generally speaking, the receptors  are not uniformly distributed spatially, which  makes the network architecture more complex. It is also well known that the intracellular space is crowded with organelles and various obstructions~\cite{Rothman1994}. The diffusing signaling molecule has to navigate between complex intracellular structures. Besides, the receptors usually  embed on the biomembrane, which is not flat. For example, the ryanodine receptor (RyR) distributes on the endoplasmic reticulum, which is folded in the cytoplasm, and some parts of different sheets may be very close to each other. The calcium signale transduction also happens at the ER-PM junction~\cite{Jing2015,Prakriya2015,Jiang2021}. In this way, signals can be communicated
between distant regions in the biomembrane. The connections
between receptors can be very complex. In the words of the network science, the signal transduction should occure on some types of the networks. However, based on the current knowledge, we do not know their explicit network architectures.  Thus, it is interesting to study the effect of  different network architectures on  signal transduction, especially whether the biological function performs stably and similarly with different network architectures.

The bistable regime can emerge from the network dynamics  in a study of the propagation of nonlinear signaling on the network~\cite{Hernandez-Hernandez2017} as in electronic systems~\cite{Braichenko2018} where the network architecture was not considered. In their work, only two states, a closed state and an open state, were considered for a receptor~\cite{Hernandez-Hernandez2017}.  In many signaling system, the receptor has more conformational states, the transition between these states may exhibit more complex patten of the signaling transduction.  The Ca$^{2+}$ induced Ca$^{2+}$ release mechanism (CICR)  is a basic mechanism of the calcium signal transduction system, which is important second messenger in the biological system, and widely involves in the regulation of cell life activities, including cell membrane permeability, cell secretion, metabolism and differentiation~\cite{Berridge2016,Clapham2007}.  In the literature~\cite{Keizer1996}, a simplified mechanism that mimics ``adaptation" of the RyR has been developed to reproduce experimental
data from cardiac cells.  In such model, the nodes of the calcium signaling network, RyRs, have four states. Moreover, among the transitions between these states, there are two fast processes and one slow process.  
In this study, we will study  the CICR on the network to explore the behavior of signaling networks with multistate in which nodes are regulated by reaction rates nonlinearly.  
In  next section, we will present  explicit CICR mechanism  and corresponding mathematical description of the states and the networks considered.  Then, the dynamics of CICR on three networks will be studied, and time revolution of  fractions of states will be simulated with different network architectures.  The pattern of steady states will be also studied to provide more informations about the possible bistable regime from the CICR mechanism.


\section{CICR Mechanism}

\begin{figure*}[h!]
\begin{center}
\includegraphics[width=75mm]{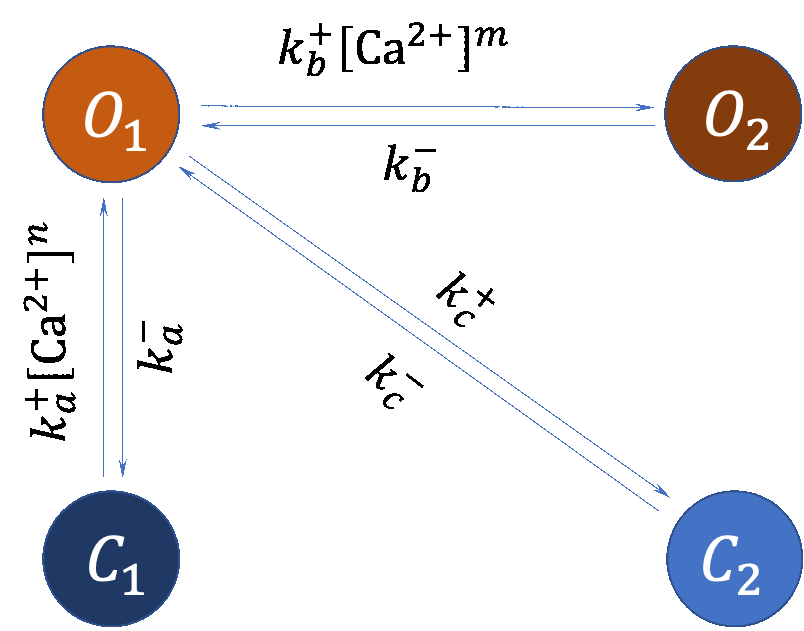}
\end{center}
\caption{Schematic diagram of transitions among the four states of the
RyR~\cite{Keizer1996}.
}
\label{Fig: 1}
\end{figure*}

In the current work, we take an RyR, the receptor of the CICR mechanism,  as a node of  network. RyR has four states as shown in Fig.~\ref{Fig: 1}. States $C_1$ and $C_2$ are closed states, in which no calcium will be released, and states $O_1$ and $O_2$ are open states, in which RyR releases calcium. 
The transitions between different states of an RyR in an environment with concentration $[{\rm Ca}^{2+}]$ are also illustrated in  Fig.~\ref{Fig: 1}.
The relevant parameters have been extracted from the experiment, and listed in  Table~\ref{Tab: constant}.
\renewcommand\tabcolsep{0.75cm}
\renewcommand{\arraystretch}{1.5}
\begin{table}[h!]
\begin{center}
\caption{RyR kinetic constants~\cite{Keizer1996}. The parameters $n$ and $m$ were also determined as 4 and 3, respectively.}
\label{Tab: constant}
\begin{tabular}{ccccccc}\bottomrule[2pt]
rate&$k^+_a$&$k^-_a$&$k^+_b$&$k^-_b$&$k^+_c$&$k^-_c$\\\hline
Value&1500&$28.8$&1500 &385.9 &1.75 &0.1\\
Unit&$\mu$M$^{-4}$s$^{-1}$&s$^{-1}$&$\mu$M$^{-3}$s$^{-1}$ &s$^{-1}$&s$^{-1}$ &s$^{-1}$\\
\toprule[2pt]
\end{tabular}
\end{center}
\end{table}

Such model was proposed in Ref.~\cite{Keizer1996} to mimic the experiment.  The states $C_1$ and $C_2$ and their conversions are the basic mechanism, and such two-state model's behavior on network was also discussed in Ref.~\cite{Hernandez-Hernandez2017}. The state  $C_1$ is dominant at a low concentration $[{\rm Ca}^{2+}]$, for example 0.1~$\mu$M.  If the $[{\rm Ca}^{2+}]$ increases, the RyR will be activated from the closed state $C_1$ to an open state $O_1$ at a rate of $k^+_a[{\rm Ca}^{2+}]^n$, and deactivates back to $C_1$ state at a rate of $k^-_a$.  Such simple two-state model is not enough to describe the experimental phenomenon.  As suggested in Ref.~\cite{Keizer1996}, to keep the plateau open probability, the open state $O_1$ also may be activated  to the second open state $O_2$  at a rate of $k^+_b[{\rm Ca}]^m$ and  back to the first open  state $O_1$ at a rate of $k^-_b$.   It is also important to obtain the bistable regime in the current work. To describe the adaption phenomenon, the transition between first open state $O_1$ and the second closed state $C_2$ is also added, which is independent on the concentration $[{\rm Ca}^{2+}]$. And compared with the rates of processes $C_1\leftrightarrows O_1$ and $O_1\leftrightarrows O_2$, the  process $O_1\leftrightarrows C_2$ should be slow, which makes the decrease of the fractions of  open states be a long-time-scale process.  With such model, the main experimental behavior can be reproduced. For simplification,  the conversions between $C_1$, $C_2$ and $O_2$  are neglected as in Ref.~\cite{Keizer1996}. 

With the above mechanism, we study the dynamics of the CICR on the network.  For a network with $N$ nodes, the state of a node can be denoted as vectors, [1, 0, 0, 0], [0, 1, 0, 0], [0, 0, 1, 0], [0, 0, 0, 1] for $C_1$, $O_1$, $C_2$ and $O_2$, respectively. The overall state of the network can be written as a $4\times N$ matrix $S(t)$, which  revolves with  time $t$.  
Besides the nodes, i.e., RyR, we need  connections between  them to study the dynamics of the  CICR network.  
As in Ref.~\cite{Hernandez-Hernandez2017}, we assume the  concentration $[{\rm Ca}^{2+}]$  at an RyR is determined by a general background concentration $c_0$ and calcium released by other RyRs. Hence, it can be obtained by the states of other nodes as 
\begin{equation}
[{\rm Ca}^{2+}](t)=c_0+r\sum^N_{mj} A_{ij}S_{mj}(t),\label{Cal}
\end{equation}
where $m=1,2,3,4$ corresponds to states $C_1$, $O_1$, $C_2$ and $O_2$, respectively, and $i$ or $j$ are the number of the nodes.
It is nature to assume a small general background concentration $c_0$, 0.1~$\mu$M,  in this work, and the  concentration is mainly from the calcium released from other nodes. Hence, only the nodes in two open states contribute to the concentration, which is denoted as $S_{2j}$ and $S_{4j}$. The $A_{ij}$ denotes adjacency matrix, which is different for different types of networks. The parameter $r$ gives the strength of connection between two nodes, which is assumed as a general constant. A more appropriate parameter can be defined as $s=r\langle k\rangle/c_0$ with the average degree  $\langle k\rangle=1/N\sum_i^N k_i$ and $k_i=\sum_i^N A_{ij}$, which will be adopted in the following calculation (its meaning can be seen more clearly under the mean-field ansatz in section~\ref{steady state}  \ \  \  ).  

\section{Dynamics of CICR on the network}

In this section, we consider the dynamics of CICR on the network. Since it is difficult to detect the explicit network architecture of the calcium signal transduction, in our study we perform the investigation with three different types of networks,  Erd\"os-R\'enyi (ER) network,  Watts-Strogatz (WS) network and BaraB\'asi-Albert (BA) network.  The networks are generated with the help of the NetworkX package in Python Language. In the ER network, the two nodes are connected with a fixed probability $p$.  In the WS network, every node connects to $k$ neighbor nodes, and two nodes are connected  with fixed probability $p$. Such network may be more appropriate to describe the behavior of the CICR mechanism because usually the calcium released by a RyR  induces the release of  neighbor RyRs  on the endoplasmic reticulum while some other distant RyRs may be also activated due to the folding of the membrane. We also consider the BA network, which is formed by adding a sequence of $m$ new nodes to an existing network. By the generated network, we have the network adjacency matrix $A$ in Eq.~(\ref{Cal}). 

With overall state matrix $S$, it is easy to characterize the activity of the network by computing the fraction of node which is in an state $i$ at time $t$ as 
\begin{align}
p_i(t)=\frac{1}{N}\sum_{j=1}^NS_{ij} (t).
\end{align}

To study the dynamics, we also should consider an ensemble of initial conditions where  activity level of  network is varied. We choose initial conditions that the node $i$ has a probability $h$ at $C_1$ and $1-h$ at $O_1$. For simplicity, we do not consider other possible initial conditions with $O_2$ and $C_2$ states, which should provide similar conclusion. Here we simulate initial conditions where $h=i/(K-1)$ and $i=0,1,...,K-1$.  With such treatment, we can explore the time revolution of a range of random initial conditions $S^i(0)$.

With the network adjacency matrix $A$ and the initial state matrix  $S^i(0)$ obtained  as above, we can simulate the time revolution of the dynamics of the CICR on the network under the mechanism  in  Fig.~\ref{Fig: 1}.  In the simulation, we adopt network with $N=500$ nodes, and choose $K=6$ random initial conditions.  We consider the time revolutions of the fractions of four states on three networks considered. The parameters are chosen as $p=0.1$ for ER network, $k=5$ and $p=0.1$ for WS network, and $m=10$ for BA network. Since we need synchronize the time for all nodes in the network to give the explicit time revolution of fractions of four states of node,  the Gillespie algorithm is not practical, so we perform this work by fixing time step to simulate the transitions between the four states of the RyR in the current work.

\begin{figure}[h!]
\begin{center}
\includegraphics[width=125mm]{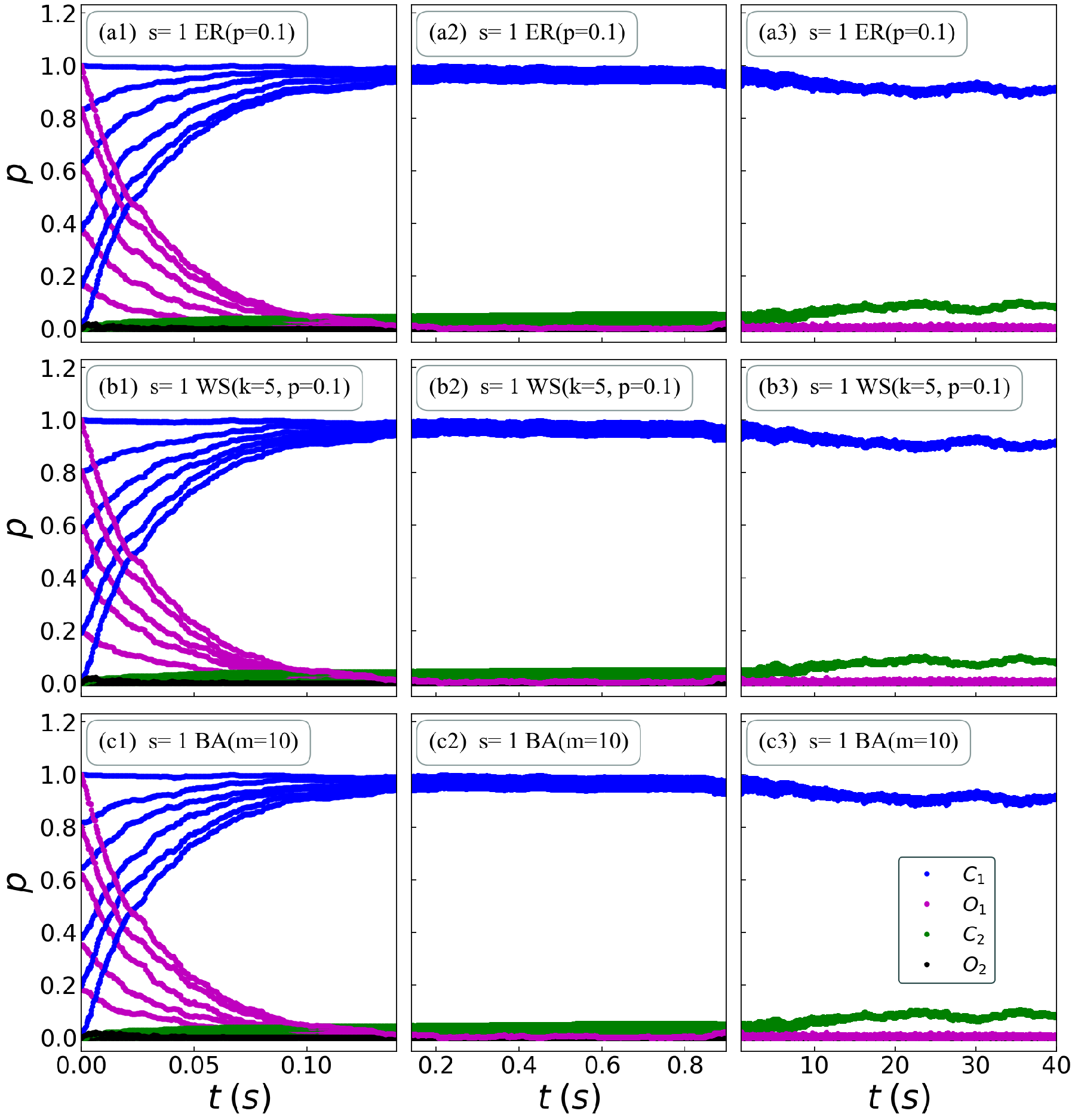}
\end{center}
\caption{The fractions of  states $C_1$ (blue dot), $O_1$ (magenta dot), $C_2$ (green dot) and $O_2$ (dark dot) on three networks as a function of time for an ensemble of initial conditions with strength of  connection $s=1$. The panels (a1, a2, a3) are for the ER networks with $p=0.1$ at short, medium, and long time scales, respectively. The panels (b1, b2, b3) are for the WS networks with $K=5$ and $p=0.1$, and  the panels (c1, c2, c3) for the BA networks with $m=10$. In the calculation, 100000 time steps are adopted.
}
\label{Fig: s1}
\end{figure}

\begin{figure}[h!]
\begin{center}
\includegraphics[width=125mm]{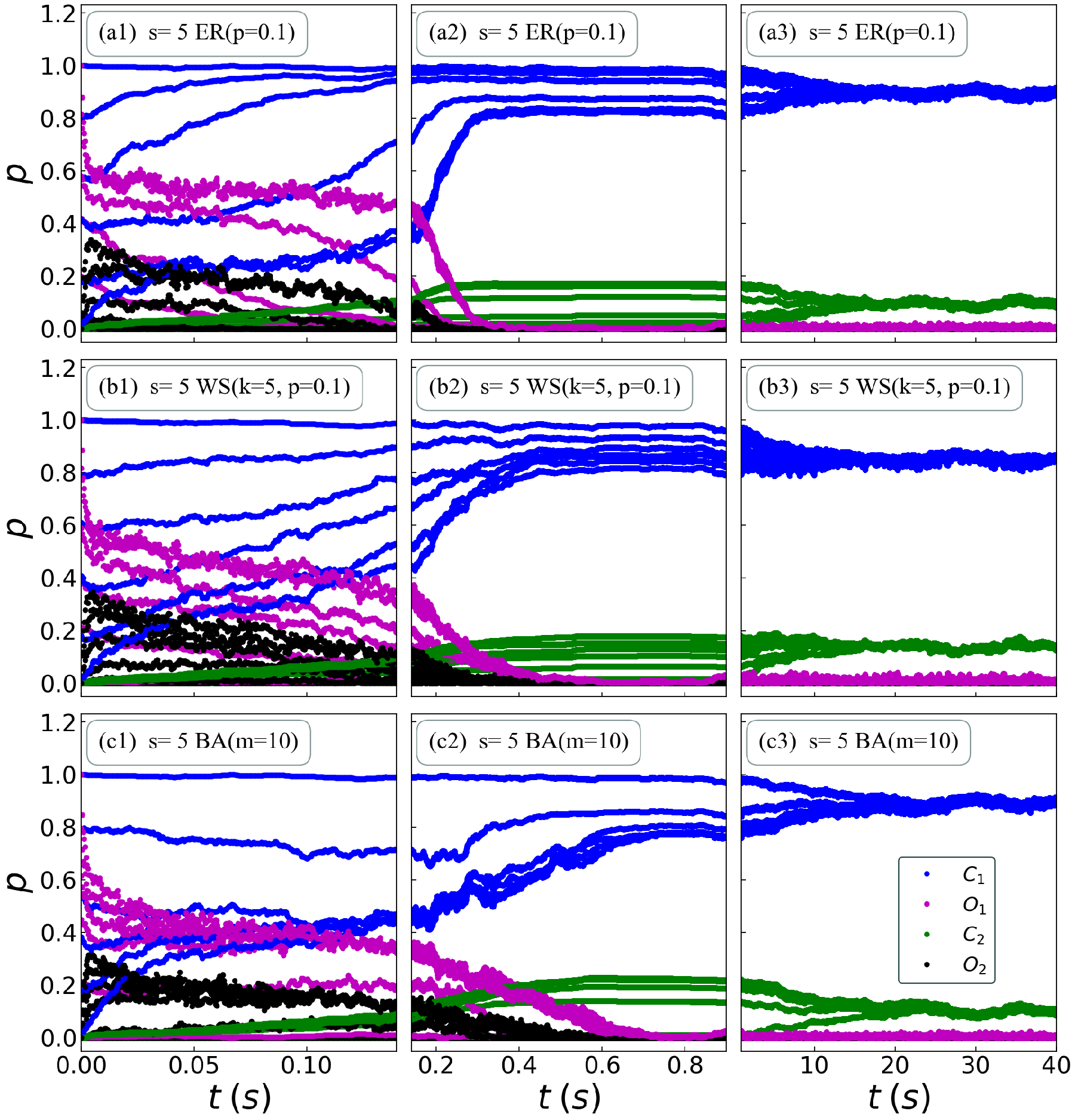}
\end{center}
\caption{The fractions of four states. The same as Fig.~\ref{Fig: s1}, but with strength of  connection $s=5$.  }
\label{Fig: s5}
\end{figure}

\begin{figure}[h!]
\begin{center}
\includegraphics[width=125mm]{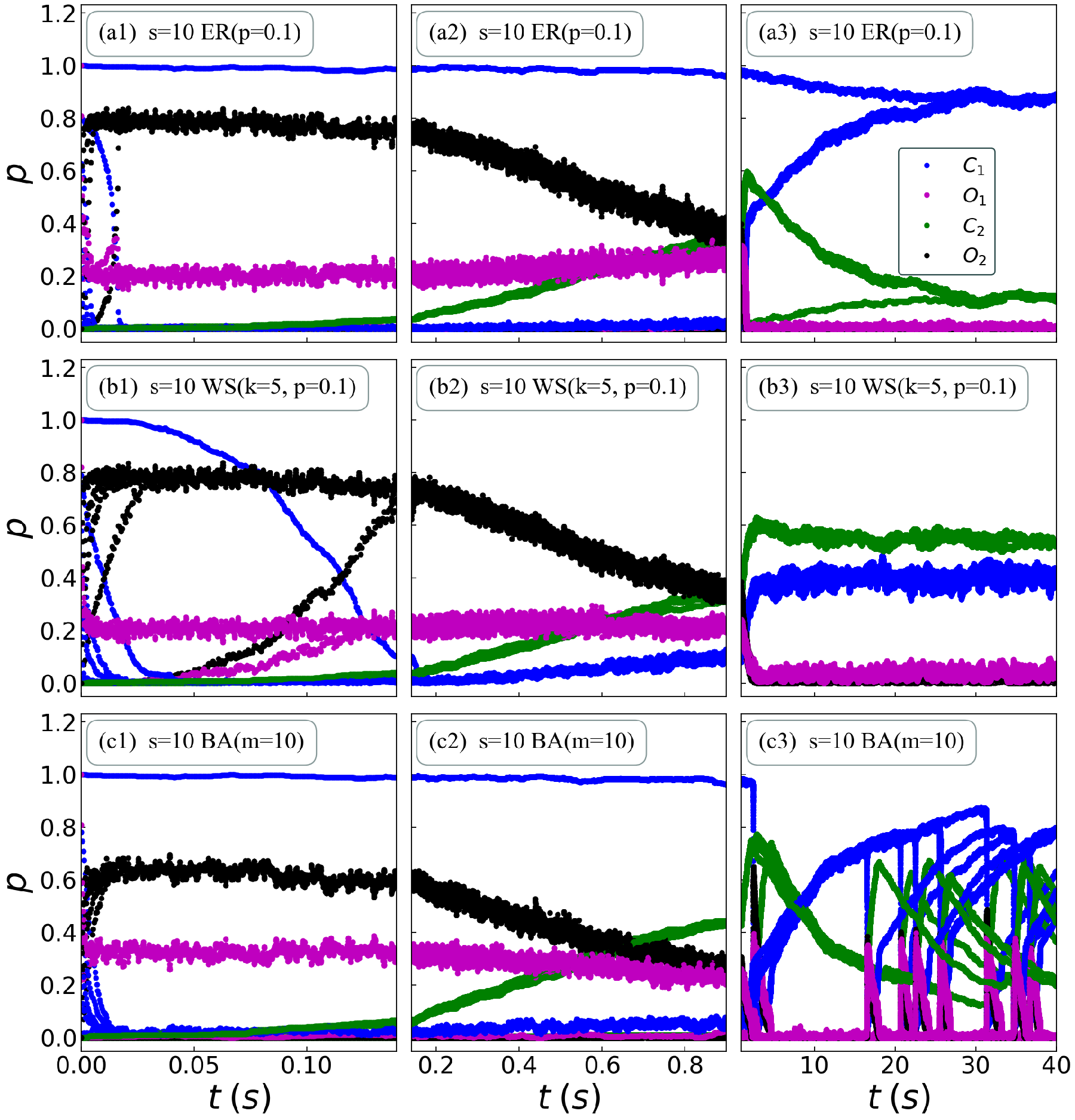}
\end{center}
\caption{The fractions of four states. The same as Fig.~\ref{Fig: s1}, but with strength of  connection  $s=10$. 
}
\label{Fig: s10}
\end{figure}

\begin{figure}[h!]
\begin{center}
\includegraphics[width=125mm]{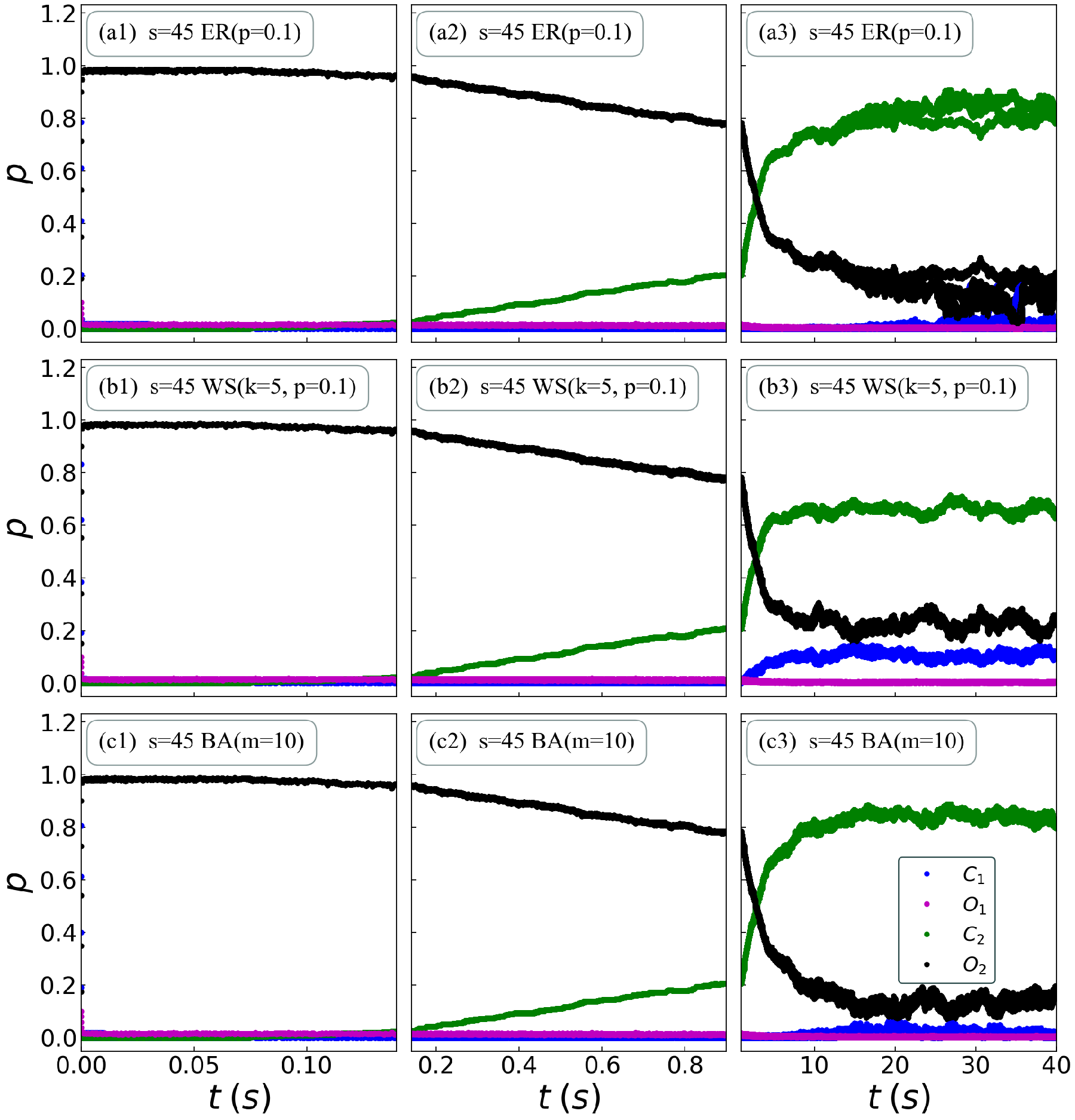}
\end{center}
\caption{The fractions of four states. The same as Fig.~\ref{Fig: s1}, but with strength of  connection  $s=45$. 
}
\label{Fig: s45}
\end{figure}

 In Ref.~\cite{Hernandez-Hernandez2017}, the nonlinear network with a closed and an open state exhibits bistable regime with a small $\eta$ in their work. If we only consider $C_1$ and $O_1$ state, an explicit calculation suggests that there are no bistable regime emerging because the values of rates $k^+_a$ and $k^-_a$ shown in  Table~\ref{Tab: constant}, which were extracted from experimental data   leads to a large $\eta$.  It is interesting to see that the bistable regime recovers after state $O_2$  is added. However, the  inclusion of the slow process $O_1\leftrightarrows C_2$  breaks such bistable regime, which can be seen in the results at different strengths of connection $s$ illustrated in Figs.~\ref{Fig: s1}-\ref{Fig: s45}. 

In Fig.~\ref{Fig: s1}, the results with small strength of connection $s=1$ are presented. We  provide the results  at both short time and long time scales to show the effects of fast processes and slow process. The results at medium time scale are also given as a reference of general picture of the time revolution. The time revolution of the fractions of four states on three networks exhibit very similar behaviors. The open state $O_1$  deactivates  within a relaxation time $\tau\sim 0.1$~s (see Fig.~\ref{Fig: s1}a1, b1, and c1). At such time scale, the state  $O_2$ is nearly not activated, and the fraction of state $C_2$ increases very slowly. Such pattern keeps up to a time about 1~s  (see  Fig.~\ref{Fig: s1}a2, b2, and c2). At the scale of 10~s , the slow process takes effect. The fractions of states $C_2$ and $C_1$ increase and decrease a little, respectively, and approach to steady states  (see Fig.~\ref{Fig: s1}a3, b3, and c3).

In Fig.~\ref{Fig: s5}, the results with  $s=5$ are presented.  Compared with the results with strength $s=$1, the behaviors of the activity of the network become more complex.  At short time scale, the state $O_1$  deactivates more rapidly than the case with $s=1$. The fast deactivation is due to the larger $s$. The larger connection between the nodes makes the propagation of the calcium signal faster, which consequently  leads to fast deactivation. However, different from the case with $s=1$, at a time about 0.02~$s$, the deactivation of $O_1$ becomes relatively slower, which means a steady state, but disturbed by the slow process $O_1\leftrightarrows C_2$. With the time revolution, the fraction of state $O_1$  approaches to zero finally as shown in the mediate panels. In fact, if we neglect the slow process, the activity of the network  enters a bistable regime at long time scale. However, the effect of the slow process $O_1\leftrightarrows C_2$ makes the time not enough to exhibit such behavior, which will be discussed later. At long time scale, the tendencies of  fractions of states $O_1$, $C_1$,  and $O_2$ at  short time scale are smeared. A stable pattern as in the case with $s=1$ emerges at time $t>$20 s.  Besides, with strength $s=5$, three networks also exhibit analogous behaviors.

In Fig.~\ref{Fig: s10}, the results with $s=10$ are presented.  
In this case, the quasibistable regime becomes more clearly.  At short time scale, the quasi-steady states can be seen clearly (here and hereafter, we will call the approximate steady state and approximate bistable regime at short time scale as ``quasi-steady state" and ``quasi-bistable regime" to distinguish  them from the real steady state at long time scale and bistable regime only with fast processes). With the increase of the strength of connection $s$, the calcium signal becomes easier to propagate from a node to another one. The relaxation time of the fast process becomes smaller, about 0.01~s with connection $s=5$. However, as shown in  Fig.~\ref{Fig: 1}, the slow process is independent of the $s$, which is the reason why the quasi-steady states here are more obvious than those with small strength.  One can observe the transitions from a quasi-steady state to another quasi-steady state. For example, on WS network, the state $O_1$ has two quasi-steady states, one is about 0 and the other about 0.2. At a time about 0.05~s, the network leaves the former and jumps to the latter. The  state $O_2$ also may jump from a sate about 0 to the state about 0.8. The values of the quasi-steady states for ER and WS networks  are almost the same. However, the BA network has a larger fraction for state $O_2$. At a time scale of 0.1~s, the fraction of state $O_2$  decreases until  reaching zero. 

The decrease of fraction of state $O_2$ can be seen at time scale of 1~s. The fraction of state $C_2$ increases first, but at long time scale its behaviors are different for different networks. On ER network, the fraction of state $C_2$ tends  toward a stable value about 0.1 and the one of state $C_1$ toward a value about 0.9. The nodes of the network are almost occupied by these two states at long time scale, the fractions of states $O_1$ and $O_2$ are about zero. On WS network, the stable value for state $C_2$ is about 0.5, higher than the one for state $C_1$, about 0.4. Besides, the system approaches steady states at time smaller than 10~s, which is different from the pattern on the ER network. For  BA network, the tendencies  of fractions of four states are analogous to the case on ER network. However, the statistical transitions  between the two steady states and state $O_1$ by fast processes  make the situation more complicated because states $C_1$ and $C_2$ connected by $O_1$, which is usually very low at long time scale. 

Now, we turn to a large strength of connection $s=45$, and the results are presented in Fig.~\ref{Fig: s45}.  With such large connection, the calcium signal propagates on the network very fast. Hence, at short time scale, one can find that after a very small relaxation time, the states of the network concentrates to the state $O_2$.  The quasi-bistable regime in previous figures disappears in the large strength. With time revolution, the fraction of  state $O_2$ decreases, and reaches a value about 0.2 or smaller at larger time scale. The state $C_2$  becomes dominant. The fraction of state $O_1$   keeps a low level at all time scales, while the state $C_1$ is observable on WS scale at long time scale. 

\section{Pattern of steady states on network}
\label{steady state}

In the above, we present the time revolution of the  fractions of four states on three networks. One can find that at small strength of connection $s$, the network is nearly not activated. With increasing of the strength of connection, quasi-bistable regime emerges at short time scale, where the fast processes play the dominant role.  However, the low process breaks the quasi-steady states with the time revolution, and network revolves to the steady states at long time scale. And with a very large strength of the connection, the quasi-bistable regime disappears. Since the steady states are more interesting in biological function.  In this section, we discuss the dependence of the quasi-steady states on networks.

To understand the basic feature of the network dynamics, we first provide the mean field (MF) ansatz of the system considering homogeneous~\cite{Hernandez-Hernandez2017}. Under this ansatz, we assume all nodes revolve with time in the same manner, which  leads to the state function $S_{ij}=S_{i0}$ with $i$ and $j$ denoting the four states and $N$ nodes, respectively. Hence the calcium concentration can be written as   
\begin{equation}
[{\rm Ca}^{2+}]=c_0+sc_0(p_{O_1}+p_{O_2}).
\end{equation}

Here we apply $\langle k\rangle=\sum^N_{j} A_{ij}$ and $S_{20}=p_{O_1}$ and $S_{40}=p_{O_2}$ under MF ansatz. Here and hereafter, for fractions of the states, we use $C_1$, $O_1$, $C_2$ and $O_2$ instead of 1, 2, 3, and 4 for convenience to understand.
The dynamics equation can be written as,
\begin{align}
\frac{dp_{C_1}}{dt}&=-k^+_a[c_0+sc_0(p_{O_1}+p_{O_2})]^4 p_{C_1}+k^-_ap_{O_1},\label{C1}\\
\frac{dp_{O_1}}{dt}&=k^+_a[c_0+sc_0(p_{O_1}+p_{O_2})]^4 p_{C_1}-k^-_ap_{O_1}\nonumber\\
&-k^+_b[c_0+sc_0(p_{O_1}+p_{O_2})]^3 p_{O_1}\nonumber\\
&+k^-_bp_{O_2}-k^+_c p_{O_1}+k^-_cp_{C_2},\label{O1}\\
\frac{dp_{O_2}}{dt}&=k^+_b[c_0+sc_0(p_{O_1}+p_{O_2})]^3 p_{O_1}-k^-_bp_{O_2},\label{O2}\\
\frac{dp_{C_2}}{dt}&=k^+_cp_{O_1}-k^-_cp_{C_2},\label{C2}
\end{align}
with $p_{C_1}+p_{O_1}+p_{C_2}+p_{O_2}=1$.

Based on the parameters in Table~\ref{Tab: constant}, the processes in Equations from (\ref{C1}) to (\ref{O2})
are fast processes while the last one is slow process. If we consider all three processes, we will reach the steady states at  long time scale. Hence, first, we consider the case with the last process $O_1\leftrightarrow C_2$ turning off. With such treatment we can obtain steady states at a long time scale which correspond to results of full model at short time scale. The steady states corresponds to the  stationary points, which satisfy the algebraic condition as 
\begin{align}
0&=-k^+_a[c_0+sc_0(p_{O_1}+p_{O_2})]^4 p_{C_1}+k^-_ap_{O_1},\nonumber\\
0&=k^+_b[c_0+sc_0(p_{O_1}+p_{O_2})]^3p_{O_1}-k^-_bp_{O_2},\label{MF1}
\end{align}
where $p_{C_1}+p_{O_1}+p_{O_2}=1$, and we still use a low background $c_0=0.1~\mu$M and consider the variation of the parameter $s$, which reflects the strength of the connections between nodes on the network. 

\begin{figure}[h!]
\begin{center}
\includegraphics[width=125mm]{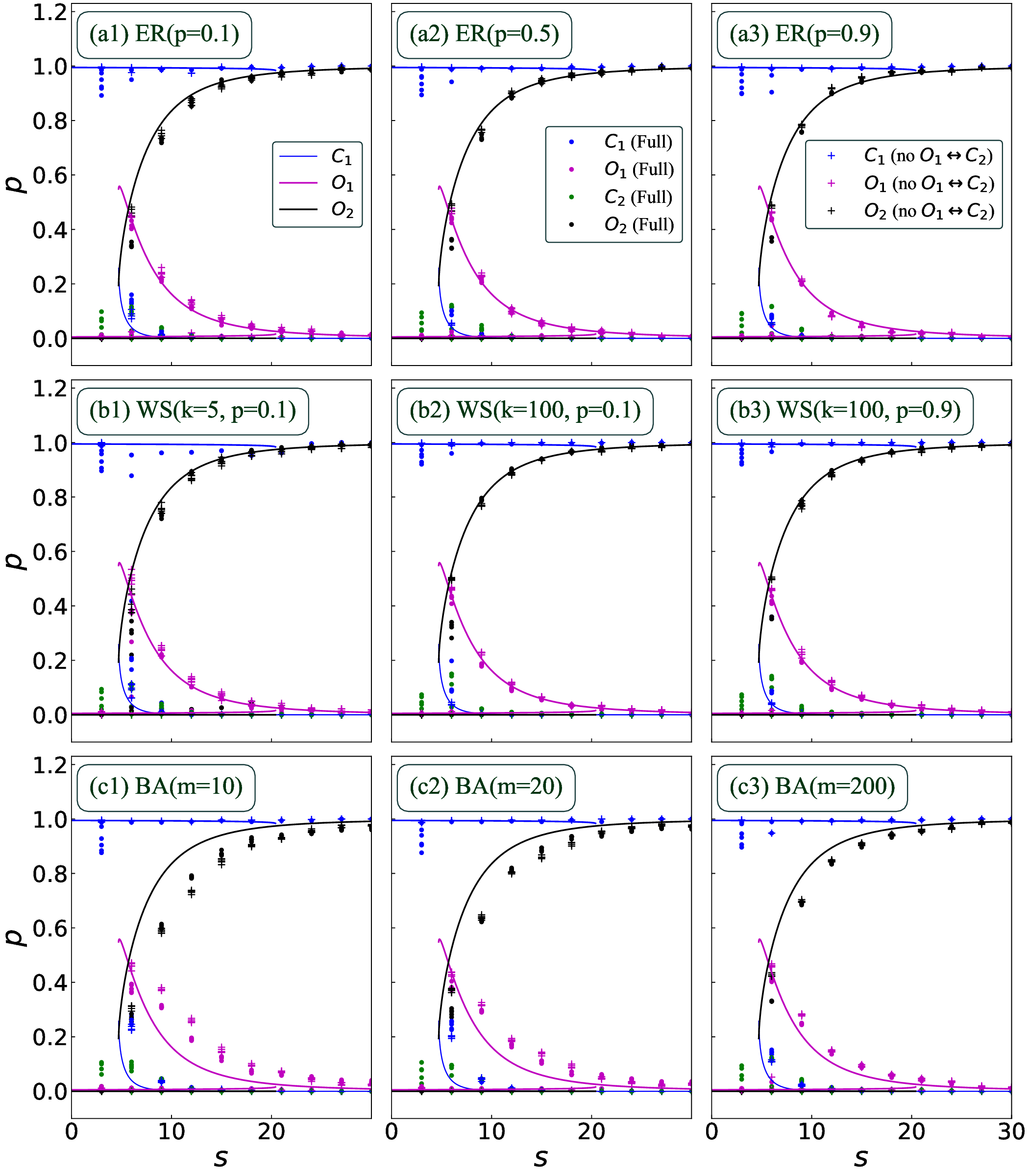}
\end{center}
\caption{The steady states on networks with different parameters with the variation of the strength of connection $s$.  The lines are for the MF ansatz with the process $O_1\leftrightarrow C_2$ turning off (Eq.~\ref{MF1}). The symbols ``$\cdot$" and ``$+$" are for the results at small time scale with full model,  and results after turning off the $O_1\leftrightarrow C_2$ process, respectively.  
}
\label{Fig: steady}
\end{figure}

In Fig.~\ref{Fig: steady}, we present the steady states obtained with above conditions. Under the MF ansatz, the difference of networks is smeared, and the effect of  network is absorbed into the strength parameter $s$.  For the results with low strength, the network can not be activated as in Fig.~\ref{Fig: s1}, which leads to a  meaningless steady state with $C_1=1$. With the strengthening of the connection between nodes, a bistable regime of fraction of state $C_1$ emerges at $s\sim5$. Besides the steady state at $C_1\sim 0$, there exists another steady state at about 0.2, which decrease rapidly with the increase of the strength $s$. The steady state at about 1 for sate $C_1$  keeps until an $s$ about 20, and disappears there. For state $O_1$, there exists a steady state at small $s$, and a new steady state about 0.5 appears at $s$ about 5 and decreases to zero slower than state $C_1$.  At an $s$ about 20, the steady state about 0   for state $O_1$ disappears and only one steady state keeps. The state $O_2$ also exhibits  bistable regime with  a steady state about 5 and a steady state about 0. The higher one increases to about 1 with the increase of $s$, and the steady state about 0 disappears at $s$ about 20 also. 
Hence, under MF ansatz, for all three states (the $C_2$ state is removed here), the fractions of  states are monostable at $s\lesssim 5$, bistable at $20\gtrsim s\gtrsim5$, and monostable again in $s\gtrsim20$.

For the realistic networks, we still consider the three networks in previous calculation with more parameters. With the $O_1\leftrightarrow C_2$ process turning off, the steady states can be reached at time long enough. As shown in Fig.~\ref{Fig: steady}, one can find that the model only with  fast processes produce a result (symbol ``+"), which fit the curves under MF ansatz  very well on the ER and WS networks. The result on BA network deviates from the MF results a little larger.  Besides, the results with different parameters of each network are similar to each other.  We also provide the results with full model (symbol ``$\cdot$"). Because the pattern of the steady states will be broken by the slow process, we choose time points $t=0.4\times 0.1^s$ to extract the values of fractions where the quasi-steady states can be seen obviously but the effect of the slow process is still very small.  It can be found that the quasi-steady states are close to these only with the fast processes at most values of $s$, except about 6. As shown in Fig.~\ref{Fig: s5}, with a  strength about 6, the effect of the slow process becomes obvious within the relaxation time of the quasi-steady states. Generally speaking, the simulation results are close to these under MF ansatz.

Now, we consider the cases with slow process, the stationary points should satisfy the algebraic conditions as
\begin{align}
0&=-k^+_a[c_0+sc_0(p_{O_1}+p_{O_2})]^4 p_{C_1}+k^-_ap_{O_1},\nonumber\\
0&=k^+_b[c_0+sc_0(p_{O_1}+p_{O_2})]^3 p_{O_1}-k^-_bp_{O_2},\nonumber\\
0&=k^+_cp_{O_1}-k^-_cp_{C_2},\label{MF2}
\end{align}
with $p_{C_1}+p_{O_1}+p_{C_2}+p_{O_2}=1$.

\begin{figure}[h!]
\begin{center}
\includegraphics[width=125mm]{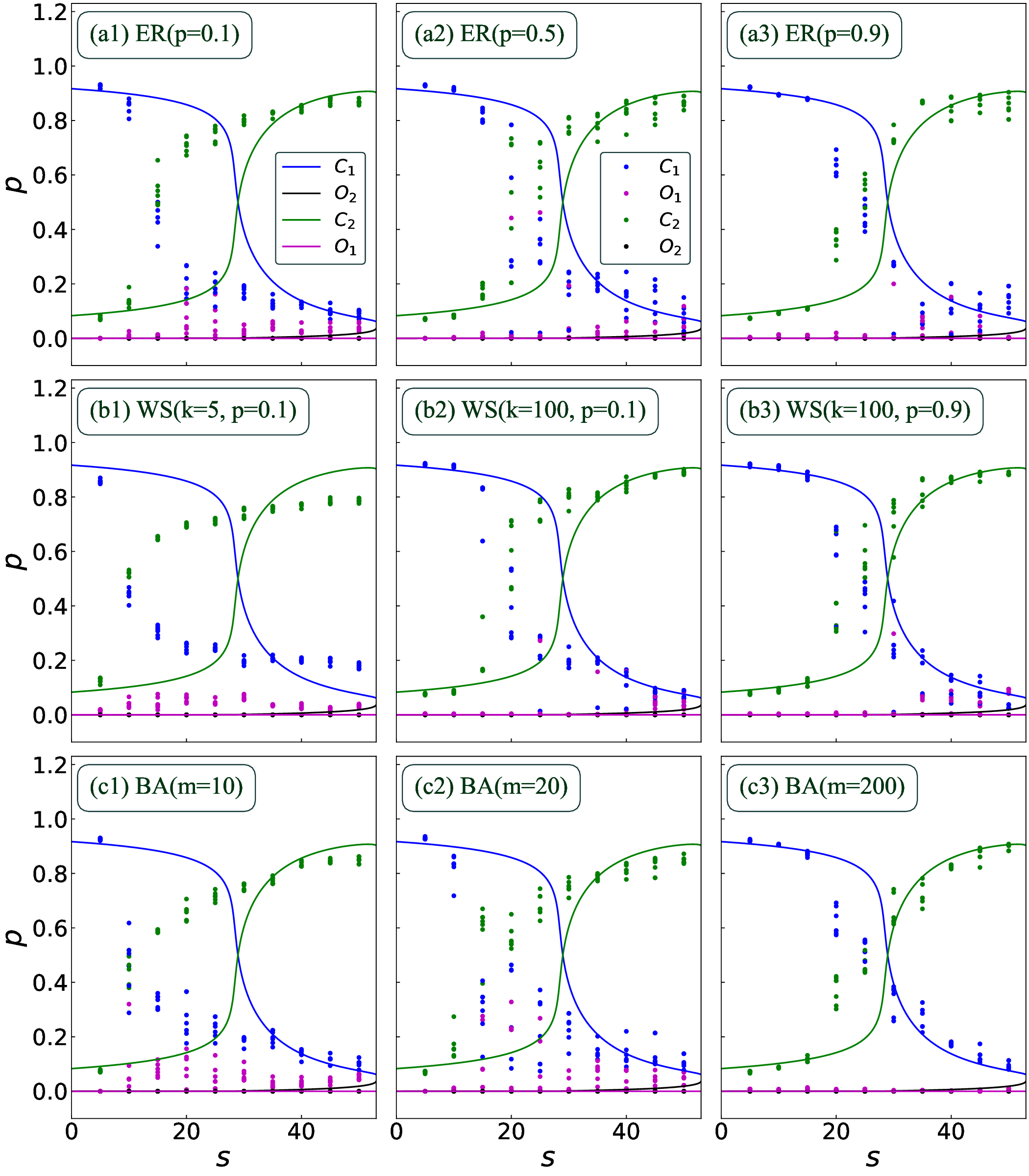}
\end{center}
\caption{The steady states on networks with different parameters with the variation of the strength of connection $s$.  The lines are for the MF ansatz  of full model (Eq.~\ref{MF2}) and the symbols ``$\cdot$" is for the results of the simulation of full model at long time scale. 
}
\label{Fig: steadyl}
\end{figure}

In Fig.~\ref{Fig: steadyl}, we present the steady states obtained from the above conditions.  At first sight,  one can find that after including the slow process, the pattern of steady states changes completely. It is also the reason of the different pattern for time revolution at short and long time scales shown in Figs.~\ref{Fig: s1}-\ref{Fig: s45}. For the $s$ from 0 to about 50, no bistable regime emerges for all four states on three networks. The MF results suggests that with weak connection, small $s$, the fraction of state $C_1$ is about 1 and decreases rapidly at about 30, and to a very small value at about 50. The fraction of state $C_2$ exhibits  almost reverse behavior, about 0.1 at $s$ about 0, and increases fast at $s$ about 30, and up to about 0.8 at $s$ about 50. The network is dominant by these two closed state, which satisfy the requirement of the biological function.  

The simulation results of the full model are also presented in the Fig.~\ref{Fig: steadyl}.  Different from results at short time scale, the results for different networks and parameters are different, and deviate from MF results obviously.  For ER network, the cross point of $C_1$ and $C_2$ is lower than MF curves.  With the increase of $p$, the simulation results become closer to the MF results. Such situations can be found also for WS and BA networks.

\section{Discussion and summary}

Signal transduction is an important mechanism in living things to regulate cell life activities. The signal usually mediates from a receptor to another receptor by releasing signaling molecular which can change the states of the receptor.  It suggests that the signal propagates on a biological network with receptors as nodes. Hence, to study the mechanism of the signal transduction, we need consider the network architecture. 

The signal transduction on the network may exhibit different behaviors from these without considering the network architecture. In Ref.~\cite{Hernandez-Hernandez2017}, the authors studied the dynamics of a two-state transition model on networks. The bistable regime was observed in their calculation.  However, if we apply their model to the important CICR mechanism in the calcium signaling transduction, the two-state transition model, states $C_1$ and $C_2$ in the current work, does not produce any bistable regime in our calculation. It is interesting to see that the bistable regime is reproduced if we include  second open state $O_2$ to states $C_1$ and $O_1$. Such results suggest that the network with more states  provides more biological patterns than in the two-states model. 

In the CICR mechanism,  the transitions between states $C_1$, $O_1$ and $O_2$ are all fast processes.  The inclusion of state $C_2$ introduces a slow process, with which the network  deactivates with time revolution. The dynamics of the CICR exhibits different patterns at different time scales. Though  with this slow process the system  tends to closed states in any case, with different strengths of the connection the system tends to different closed states. Weak connection between nodes  leads to a large fraction of $C_1$ state while strong connection to a large fraction of $C_2$ state. 

In this study we consider three types of networks, Erd\"os-R\'enyi network,  Watts-Strogatz network and BaraB\'asi-Albert network, with different parameters. The pattern at short time scale is not sensitive to the network architecture, and close to MF results. At long time scale,  only monostable regime is observed. The difference of network architectures affects the results more seriously, and  deviates from the MF results at most cases. 

In summary, our study shows that nonlinear  network  with multistate exhibits rich patterns, especially after the processes with different time scales are included. The second open state of CICR is found essential to reproduce the bistable regime at short time scale.   Our finding identifies features of biological signaling networks that may underlie their biological function. The results are also helpful to understand other polymorphic networks.   Based on the model in the current work, more complex phenomenon of calcium signal, such as intracellular calcium concentration oscillates, can be investigated in future study after including more factors, such as leak and the SERCA pump for a closed cell,  and additional PMCA pump and Ca$^{2+}$ influx for open cell. With abundant experimental information, the mechanism of the calcium signal transduction on the network can be further understood.
Vise versa, the further explicit analysis of experimental data may be helpful to unveil explicit network architecture of the calcium signal transduction.

\end{document}